\documentclass[runningheads]{svmult}

\usepackage{makeidx}   % allows index generation
\usepackage{graphicx}  % standard LaTeX graphics tool
\usepackage{subeqnar}  % subnumbers individual equations
\usepackage{multicol}  % used for the two-column index
\usepackage{physprbb}  % modified textarea for proceedings,
\makeindex             % used for the subject index
\newcommand{\msun}{\ensuremath{M_{\odot}}}
\newcommand{\zsun}{\ensuremath{Z_{\odot}}}

\newcommand{\hb}{\ensuremath{{\textrm{H}}\beta}}
\newcommand{\hda}{\ensuremath{{\textrm{H}}\delta_{A}}}
\newcommand{\hdf}{\ensuremath{{\textrm{H}}\delta_{F}}}
\newcommand{\hga}{\ensuremath{{\textrm{H}}\gamma_{A}}}
\newcommand{\hgf}{\ensuremath{{\textrm{H}}\gamma_{F}}}
\newcommand{\mg}{\ensuremath{{\textrm{Mg}}_2}}

\newcommand{\teff}{\ensuremath{T_{\rm eff}}}
\newcommand{\afe}{\ensuremath{[\alpha/{\rm Fe}]}}
\newcommand{\aca}{\ensuremath{[\alpha/{\rm Ca}]}}
\newcommand{\an}{\ensuremath{[\alpha/{\rm N}]}}
\newcommand{\feh}{\ensuremath[{{\rm Fe}/{\rm H}}]}
\newcommand{\zh}{\ensuremath[{{\rm Z}/{\rm H}}]}

\newcommand{\gapprox}{\,\rlap{\lower 2.5pt % > ungefaehr =
\hbox{$\sim$}}\raise 1.5pt\hbox{$>$}\,}
\newcommand{\lapprox}{\,\rlap{\lower 2.5pt % < ungefaehr =
\hbox{$\sim$}}\raise 1.5pt\hbox{$<$}\,}

\begin{document}
\title*{Stellar Population Models}
\toctitle{Focusing of a Parallel Beam to Form a Point
\protect\newline in the Particle Deflection Plane}
% allows explicit linebreak for the table of content
%
\titlerunning{Stellar Population Models}
\author{Claudia Maraston}
%\inst{1}}
%
\authorrunning{Claudia Maraston}

\institute{Max-Planck-Institut f\"ur extraterrestrische
Physik, Garching, Germany}

\maketitle              % typesets the title of the contribution

\begin{abstract}
This review deals with stellar population models computed by means of
the {\it evolutionary synthesis} technique that was pioneered by
Beatrice Tinsley roughly three decades ago. The focus is on the
simplest models, the so called {\it Simple Stellar Populations}, that
describe instantaneous generations of single stars with the same
chemical composition and age. The development of these models until
very recent results is discussed, pinpointing the model uncertainties
that have been solved and those that still demand a cautionary use of
the models.  The fundamental step of calibrating the models with {\it
galactic} globular clusters, for which ages and element abundances are
known independently, is illustrated by means of key examples.
\end{abstract}

\section{Introduction} 

The evolutionary population synthesis (EPS) is the technique to model
the spectrophotometric properties of stellar populations, that uses
the knowledge of stellar evolution. This approach was pioneered by
B.~Tinsley (see Section~3.1) in a series of fundamental papers, that
provide the basic concepts still used in present-day models. The
target of EPS models are those stellar systems that cannot be resolved
into single stars, like galaxies and extra-galactic globular
clusters. The comparison with the models aims at providing clues on
the ages and element abundances of these unresolved stellar
populations, in order to constrain their formation processes, and
finds ubiquitous astrophysical uses.  The simplest flavour of an EPS
model, called {\it Simple Stellar Population} (hereafter SSP), assumes
that all stars are coeval and share the same chemical composition. The
advantage of dealing with SSPs is twofold.  First, SSPs can be
compared directly with globular cluster (hereafter GC) data, since
these are the ``simplest'' stellar populations in nature. This offers
the advantage of {\it calibrating} the SSPs with those GCs for which
ages and element abundances are independently known
(\cite{RenFu}). This step is crucial to fix the parameters that are
used to describe that part of the model ``input physics'' that cannot
be derived from first principle (convection, mass loss, mixing).  The
calibrated models can be applied with more confidence to the study of
extragalactic stellar population. Second, complex stellar systems
which are made up by various stellar generations are modelled by
convolving SSPs with the adopted star formation history
(e.g. \cite{AY86}, \cite{RocGui}, \cite{Vaz96}, \cite{KA},
\cite{BP97}), therefore the deep knowledge of the building blocks of
complex models is very important.

The article starts with a description of stellar population models in
terms of ingredients, assumptions and computational technique
(Section~2), which is followed by a historical overview of the
evolution of these models until the most recent results (Section 3).
\section{Model Structure}

\subsection{Ingredients}
The basic ingredients of a stellar population model are the stellar
evolutionary tracks and the stellar model atmospheres.
\begin{figure}[ht]
\includegraphics[width=.49\textwidth]{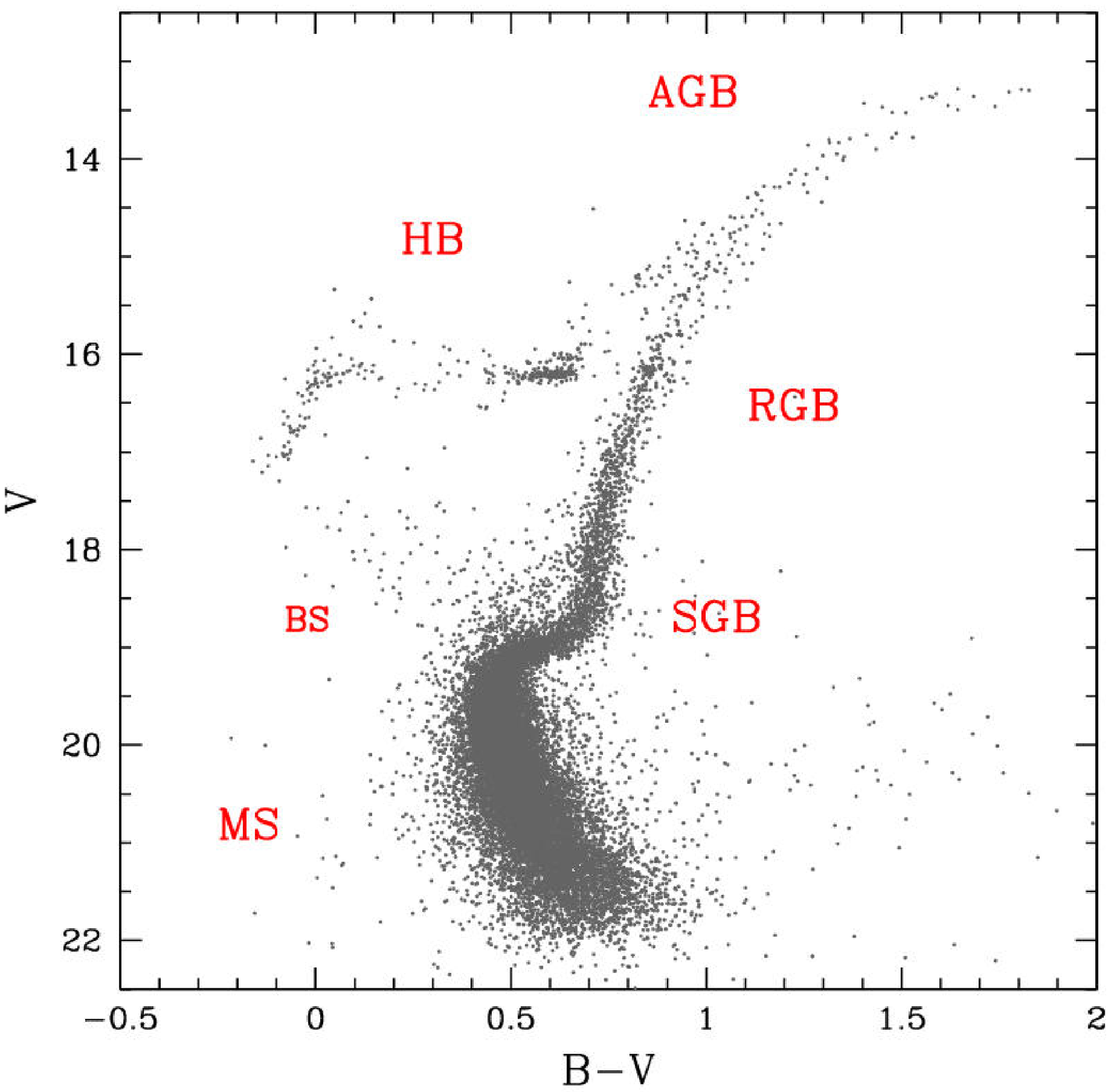}
\includegraphics[width=.49\textwidth]{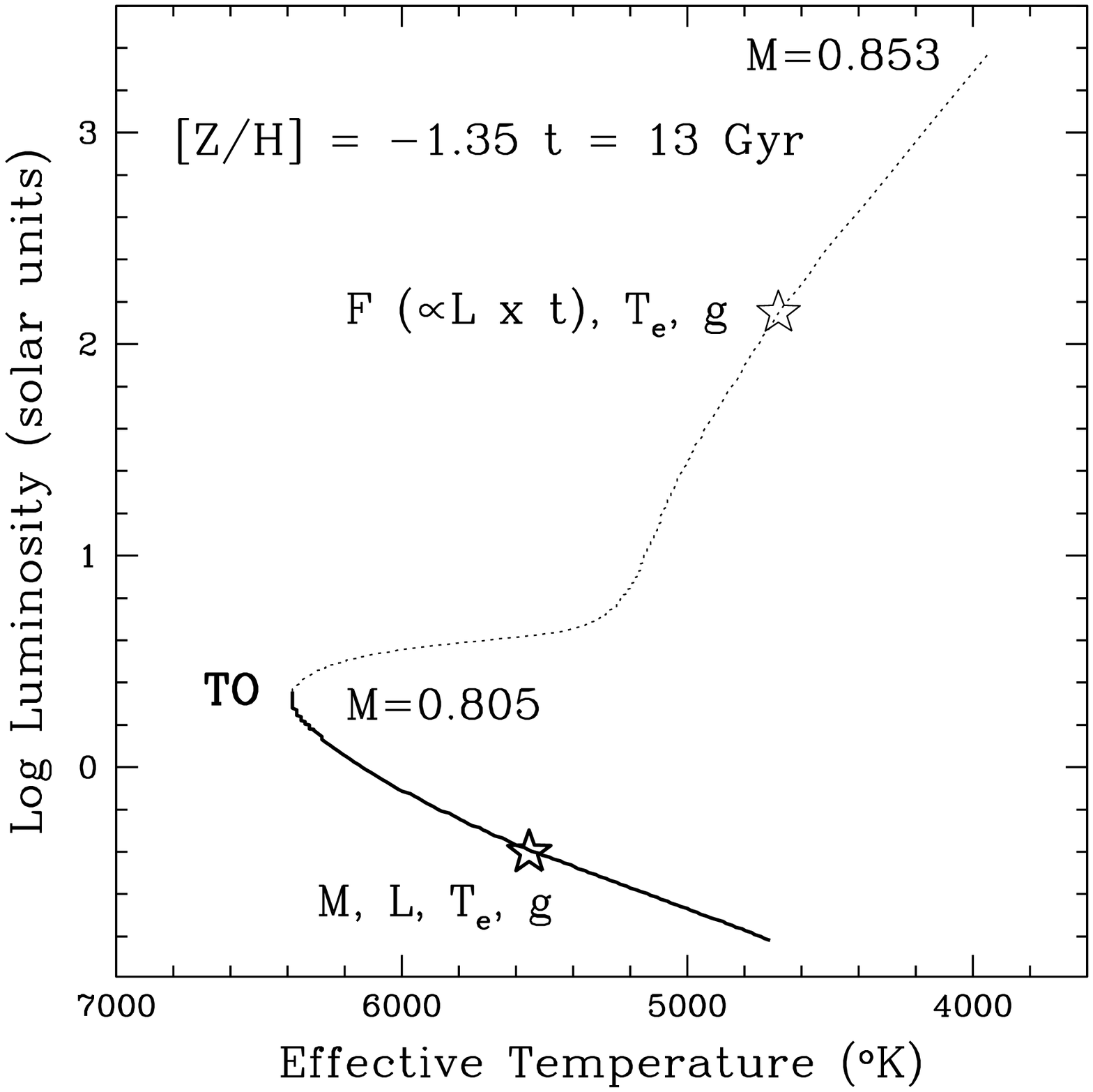}
\caption[]{{\it Left-hand panel} Observed color magnitude diagram
(CMD) of the old, metal-poor galactic GC NGC~1851 (data from
\cite{P02}). Evolutionary phases are labelled, i.e. Main Sequence
(MS); Sub Giant Branch (SGB); Red Giant Branch (RGB); Horizontal
Branch (HB); Asymptotic Giant Branch (AGB). BS mean blue
stragglers. {\it Right-hand panel}. Theoretical H-R diagram of an old,
metal-poor SSP, up to the RGB-tip (isochrone from
\cite{Casetal}). Solid and dotted linestyles mark Main Sequence and
post Main Sequence, respectively. Stellar masses at the turnoff (TO)
and RGB-tip are given.}
\label{cmd}
\end{figure}
The former trace the evolution of stars of given mass and chemical
composition through the various evolutionary phases
(Figure~\ref{cmd}), providing the basic stellar parameters -
bolometric luminosities $L$; effective temperatures \teff; surface
gravities $g$ - as functions of evolutionary timescales. The model
atmospheres describe the emergent flux as a function of these
parameters, which allows the computation of the stellar spectral
energy distribution (SED).

Various sets of stellar tracks exist in the literature, which can be
divided into two main groups according to whether the {\it convective
overshooting} is accounted for or not. Examples of overshooting tracks
are the Padova (e.g. \cite{fagoetal}), the Geneva (\cite{MM}) and the
Yale tracks (see Yi, this volume). Stellar tracks which do not include
overshooting are those by e.g. \cite{Van} and \cite{Casetal}. The actual
size of overshooting is still a matter of debate among the various
groups, but recent works favour small values (see Yi, this volume).
\begin{figure}[ht]
\begin{center}
\includegraphics[width=.49\textwidth]{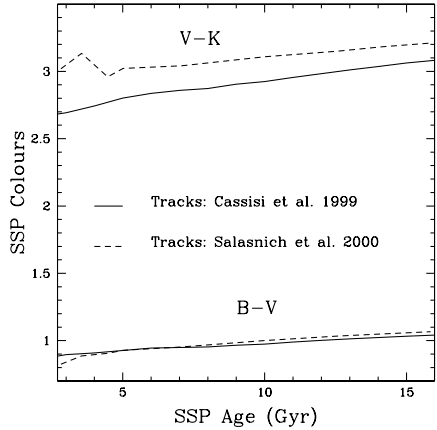}
\caption{Impact of the temperature of the RGB as described by
different tracks, on SSP colors.}
\end{center}
\label{rgb}
\end{figure}
The temperature of the Red Giant Branch is the most discrepant
quantity among the stellar tracks, that results in the largest
differences among the SSP output. This is shown in Figure~2 that
compares the B$-$V and V$-$K colors of SSP models with solar
metallicity and various ages computed with two different sets of
stellar tracks, and the same EPS.  While the B$-$V colors are
virtually identical, indicating a consistent modeling of the MS, the
V$-$K colours that depend strongly on the RGB, are very different. A
measured V$-$K$\sim~3$ is consistent with ages of $\sim 5~{\rm
Gyr}$~or~$\sim~13~{\rm Gyr}$~if the tracks by \cite{Sala00} or by
\cite{Casetal} are used, respectively. This agrees with the results by
\cite{CWB}, that attribute the large discrepancies in the V$-$K of
their respective models almost entirely to differences in the stellar
evolution prescriptions.  The temperature of the RGB is determined by
the mixing-length parameter, which regulates the stellar radius, hence
effective temperatures. In both sets of stellar tracks the
mixing-length is calibrated with the solar standard model. The reason
of such a discrepancy deserves further investigation.

The most complete grid of model atmospheres is that by \cite{K79} (and
revisions), which covers $\teff>3500~K$ for a wide range of
metallicities and gravities, the atmospheres for cooler stars being
provided by other groups (e.g. \cite{Besetal}). EPS modellers were
forced to make a collage between the various libraries, which could
introduce discrepancies in the model results. A recent improvement is
due to the works of the Basel group (e.g. \cite{LCB97}, \cite{Wes}),
that linked the various available libraries with the Kurucz`s one, and
compared the derived temperature-color relations with empirical ones,
when available. The resulting library covers virtually the whole
parameter space (\teff; $g$; \feh) required by theoretical stellar
tracks, with the exception of AGB M-type and Carbon stars (see
Section~3.6).  The spectral resolution of the model atmospheres range
between 10 and 20 \AA~.

\subsection{Assumptions}

Model assumptions are: 1) The stellar initial mass function (IMF),
that provides the mass spectrum of the population. It is common to
express IMFs as declining power-laws with single or multiple slopes,
as it follows from empirical stellar counts.  2) The helium enrichment
law $\Delta Y / \Delta Z$. 3) The amount of stellar mass-loss.
Mass-loss is a very critical assumption that impacts on the typical
observables of stellar populations, like colours and
luminosities. Mass-loss occurs in massive stars already on the Main
Sequence. In intermediate mass stars ($2<M/\msun<5$) it is most
efficient at the Asymptotic Giant Branch (AGB) and determines the fuel
and extension of the Thermally Pulsing AGB phase (herefater
TP-AGB). In low-mass stars mass loss acts already along the Red Giant
Branch (RGB) and determines the morphology of the subsequent
Horizontal Branch (HB) phase. Since the amount of mass loss cannot be
{\it predicted} by stellar evolution it must be {\it calibrated} with
globular clusters data.

\subsection{Computational procedures}
The integrated properties of SSP models, e.g. the SED, broad-band
luminosities, magnitudes and colors, are obtained from the integration
of the contributions by individual stars. Two methods in the
literature can be distinguished by the choice of the integration
variable for the stars in the post Main Sequence phase: isochrone
synthesis and fuel consumption-based methods. The right-hand panel of
Figure~\ref{cmd} shows in the theoretical H-R diagram the isochrone of
a 13 Gyr old metal-poor SSP up to the RGB-tip. The Main Sequence is
populated by stars spanning a large mass range, from
$\sim$~0.1~\msun~to the turnoff mass. The stellar luminosities are
proportional to a high power of the stellar masses, therefore the
integrated luminosity of the Main Sequence is obtained with an
integration by mass of the stellar luminosities, convolved with the
IMF. In post Main Sequence phases the mass range spanned by living
stars is very small (cf. labels in Figure~1), and the luminosity is no
longer related primarily to the mass. For the post Main Sequence two
choices are possible. 1) The evolutionary lifetime of one single mass
is adopted as integration variable. In this case the synthesis is
based on the {\it fuel consumption theorem} (see Section 3.3).  2) The
integration by mass is also performed in post-Main Sequence, in which
case the technique is called {\it isochrone synthesis} (see Section
3.4).  Pros and cons of the two methods have been longely discussed in
the literature (see e.g. \cite{CB91}, \cite{Ren94}). A recent
viewpoint will be given elsewhere.

\section{Historical overview}
The past three decades have witnessed great progress in the modelling
of stellar populations. In the next paragraphs I will describe the
basic steps that have featured the development of these models to
present-day releases, by grouping the literature works into
decades. The outstanding developments in each decade are emphasized.

\subsection{The 60's: the first color/time evolution}
A first example of stellar population model is that of \cite{CH}. The
authors describe the evolution of the B$-$V colour of the old open
cluster M67. This is accomplished with a rudimental isochrone obtained
by best fitting the observed B$-$V vs.~$M_{\rm V}$~diagram, then aged
and rejuvenated by 5 Gyr, by applying homology arguments and keeping
fixed the RGB tip in the isochrones (cf. their Fig~1). Although
stellar evolutionary computations have shown later that isochrones are
not homologous, yet the paper is important because it recognizes that
the integrated colours of stellar populations because of their
sensitivity to age, can be used to date extragalactic systems.

Following the developments of the first wide sets of evolutionary
tracks, \cite{Tin68} compute the first ``galaxy'' model. The evolution
of colour indices from $U$ to $L$ is provided, using the empirical
\teff-colour relations by \cite{J66}. Stellar tracks describe only the
upper MS and SGB, the RGB is added empirically as well as the lower
Main Sequence. The basic learnings are: i) stars must be set in
numbers $\propto$~lifetimes; ii) the post-main sequence evolution is
so fast that mass dispersion is not important (see Section~3.3).

\subsection{The 70's: the Tinsley legacy}

The 70's see the full development of B. Tinsley's fundamental work. In
a series of papers (e.g. \cite{Tin72}, \cite{Tin73}) the evolutionary
population synthesis as a method to compute the spectrophotometric
properties of galaxies is defined. Analytical approximations are
provided for the main parameters., e.g. star formation rates, IMFs,
chemical evolution, that are still used in nowadays models. The
Tinsley's models are targeted to galaxies, and a wider discussion on
them goes beyond the aim of this review. Relevant to our context is
the early modeling of near infrared colours (\cite{TinGu}), that made
evident the problem of the accuracy of the integration along the
RGB. By citing the authors {\it ``.. very slight departures from equal
spacing in the stellar lifetimes lead to unacceptable irregularities
in color because of short-lived but energetically important points''}.

\subsection{The 80's: the fuel consumption theorem}

The culmination of the efforts of the 70's is the last work of
B. Tinsley (\cite{GST}). The authors adopted the first release of the
Yale isochrones, to which empirical RGB, AGB and lower-MS were
added. The application of these models to ellipticals suggested rather
young ages, of the order of a few Gyrs. Indeed, it turned out (see
Figure~1 in \cite{Ren86}) that those Yale isochrones had an {\it
uncalibrated} mixing-length and were found not to be able to reproduce
the MS of galactic GCs. The danger at deriving ages and metallicities
of unresolved stellar populations with models based on uncalibrated
stellar ingredients motivated \cite{Ren86} to re-discuss the concept
of stellar population model. First, simple models should be preferred
against the more complicated realizations, for these models can be
straighforwardly compared to globular clusters. By {\it simple} it is
meant that all stars are coeval and chemically homogeneous, that is
the definition of SSP \`a la \cite{Ren81}. Second, two conditions
should be fulfilled: i) The use of isochrones with {\it calibrated}
mixing-length.  ii) The account of {\it every} evolutionary phase with
its {\it proper} energetic contribution. Indeed in the models of the
'70's major evolutionary phases (e.g. HB, AGB, Post-AGB) were not
included or their contribution not properly evaluated. The {\it fuel
consumption theorem} was used to show how much they can contribute. It
defines the energy conservation law for the stellar case, and is
stated as: {\it The contribution by any Post Main Sequence
evolutionary phase to the total luminosity of a simple stellar
population is proportional to the amount of nuclear fuel burned in
that phase} (\cite{Ren81}).
\begin{figure}[ht]
\begin{center}%
\includegraphics[width=.49\textwidth]{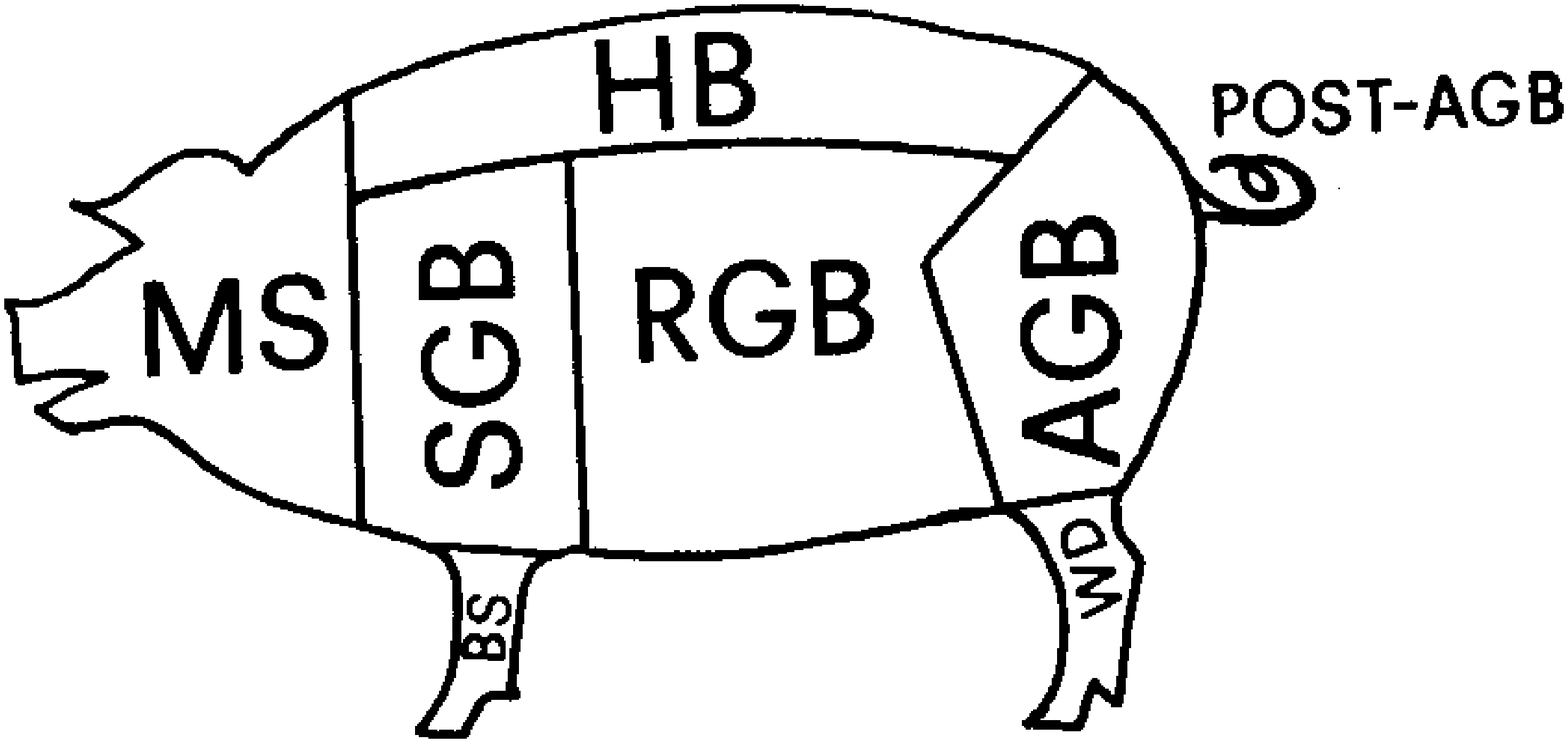}
\includegraphics[width=.49\textwidth]{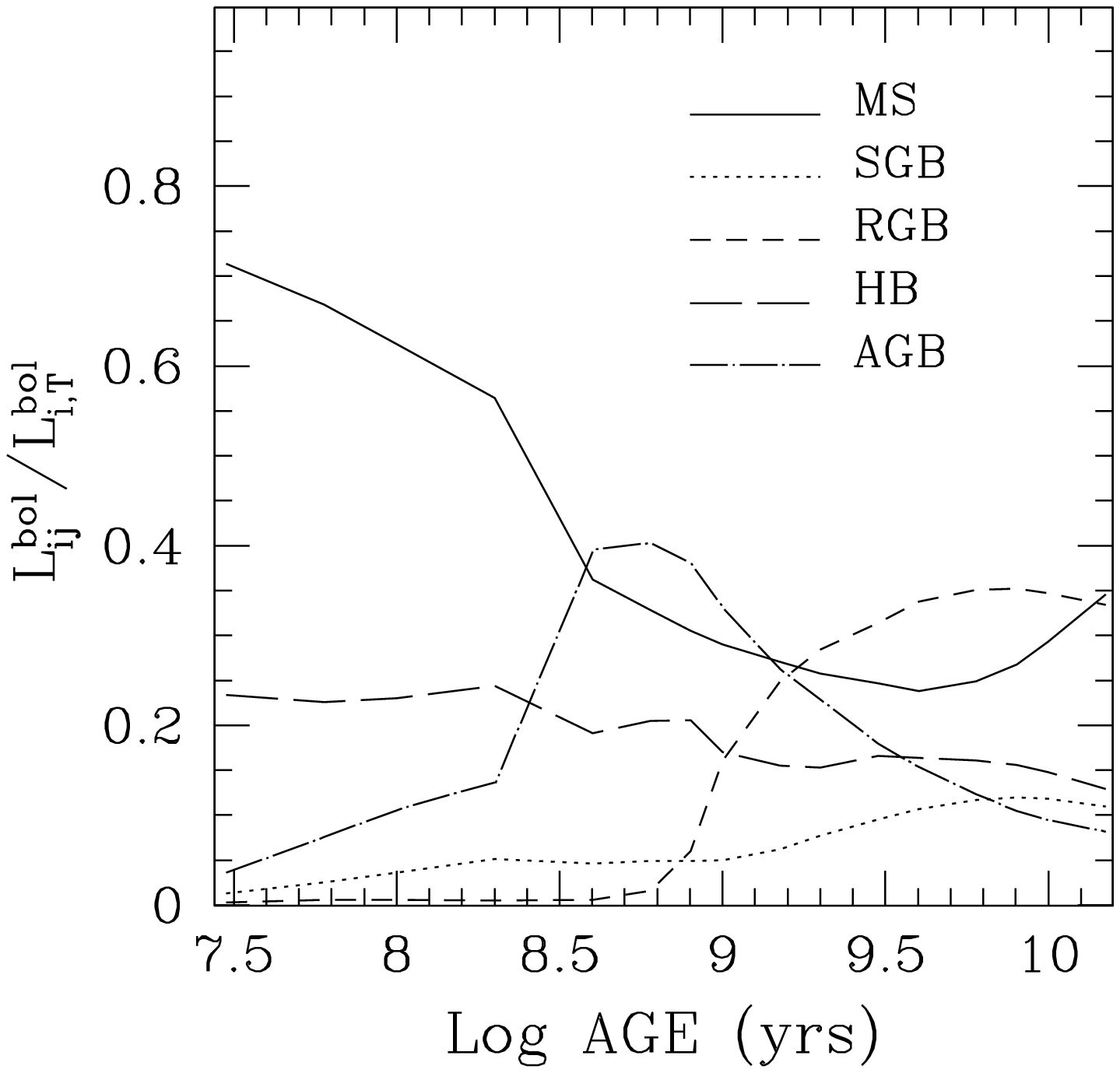}
\end{center}
\caption[]{An art view of the bolometric budget of a stellar
population (left; from \cite{Ren86}) and its quantitative evaluation
(right; from \cite{io98}). Evolutionary phases are labelled as in
Figure~\ref{cmd}.}
\label{fct}
\end{figure}
The bolometric budget of a SSP as a function of age, computed with the
fuel consumption theorem is shown in Figure~\ref{fct}. The bolometric
modeling of \cite{RB86} is extended to the monochromatic by
\cite{Buzz89}. These models span a wide metallicity range, consider
all stellar evolutionary phases and take into account various
morphologies of the Horizontal Branch. However these models are
computed only for old ages. \cite{io98} extends the fuel consumption
approach to a wider age range (Section~3.6).

Worth noting in the same decade are the first stellar population
models aimed at describing the redshift evolution of galaxies
(\cite{B83}).

\subsection{The 90's: isochrone synthesis}

The technique called {\it isochrone synthesis} is by far the most
popular method to compute EPS models, due to its straightforwardness
of definition (see Section~2.3). The isochrone synthesis boosted with
the works of Bruzual \& Charlot (but see also \cite{CBB88}). Early
implementations of the method by \cite{B83} suffered from erratic
color jumps due to poor interpolation algorythms, which were later
refined (\cite{CB91} and \cite{BC93}). Also different from the past is
the use of a nearly homogeneous set of stellar ingredients, and the
semi-empirical inclusion of the TP-AGB phase (see Section~3.6).
Although only for solar metallicity, these models span a wide age
range including for the first time very young ($t<1$~Gyr) ages. A
wider metallicity range has been provided by successive releases.
These works have defined a standard technique for most models in the
literature, which are constructed by applying isochrone synthesis to
the isochrones released by the Geneva or by the Padova group
(e.g. \cite{Tan96}, \cite{Starburst99}, latest Worthey models, latest
Vazdekis models).

\subsection{The 90's: the Lick indices}

Parallel to the diffusion of the isochrone synthesis, the 90's see the
full development of the so-called {\it comprehensive} models, the
feature of which is to provide the largest number of model output,
e.g. broad-band colours, spectral energy distributions, mass-to-light
ratios, spectral indices, surface brightness fluctuations, for the
largest number of model parameters (metallicities, ages, IMFs,
etc.). The most complete of these models for the study of {\it old}
stellar populations are the models by \cite{Wor94}. With these a
modern quantification of the age/metallicity degeneracy
(e.g. \cite{F72}) has been made, which is known as the ``3/2 rule''
$\Delta {\rm \log t}/\Delta {\rm \log Z}\sim 3/2$. But perhaps the
most important feature of the Worthey's models is to provide as first
the SSP values of the whole set of the absorption line indices in the
Lick system, the so-called Lick indices (\mg, Fe5270, \hb,
etc.). These are obtained by inserting in the Worthey's EPS code the
{\it fitting functions} (\cite{Woretal94}), that provide the stellar
index as a function of \teff, gravity and metallicity, and are
constructed with real stars. Similar fitting functions for the indices
\mg, Fe5270, Fe5335 and \hb~are provided by \cite{Buzzetal92} and
\cite{Buzzetal94}, with the corresponding SSP models in the framework
of the evolutionary synthesis of \cite{Buzz89}. When coupled with the
same evolutionary synthesis these two sets of fitting functions
produce consistent SSP indices (\cite{ioetal02}). Both fitting
functions by Buzzoni and Worthey do not take element abundance ratios
into account but depend only on total metallicity. However, the
specific abundances of given elements likely affect the absorption
features, and it is known that, e.g. the magnesium-to-iron abundance
ratio in Milky Way stars varies as a function of total metallicity. A
step forward was made by \cite{Bor95}, whose fitting function for
\mg~contains the explicit dependence of the Mg abundance relative to
Fe. However this approach has not been extended to other indices. Only
very recently SSP models of all Lick indices in which the abundances
of the specific elements are a model parameter, have been made
available (\cite{TMB02}; see Section 3.6).
\subsection{Around the turn of the century}

\subsubsection{Back to the fuel for TP-AGB}
The TP-AGB is perhaps the most critical stellar phase to be accounted
for in a SSP model, because its energetic and extension are affected
by mass-loss and nuclear burning in the envelope, both phenomena
requiring parametrizations to be calibrated with data. However, the
TP-AGB phase is the most important phase in intermediate-age
($0.2~{\rm Gyr}\lapprox t\lapprox~1~{\rm Gyr}$)~stellar populations,
contributing $\sim 40\%$ to the bolometric light (\cite{RB86}). The
fuel consumption approach allows one to include the TP-AGB phase in a
SSP model in a semi-empirical way.
\begin{figure}[ht]
\includegraphics[width=.49\textwidth]{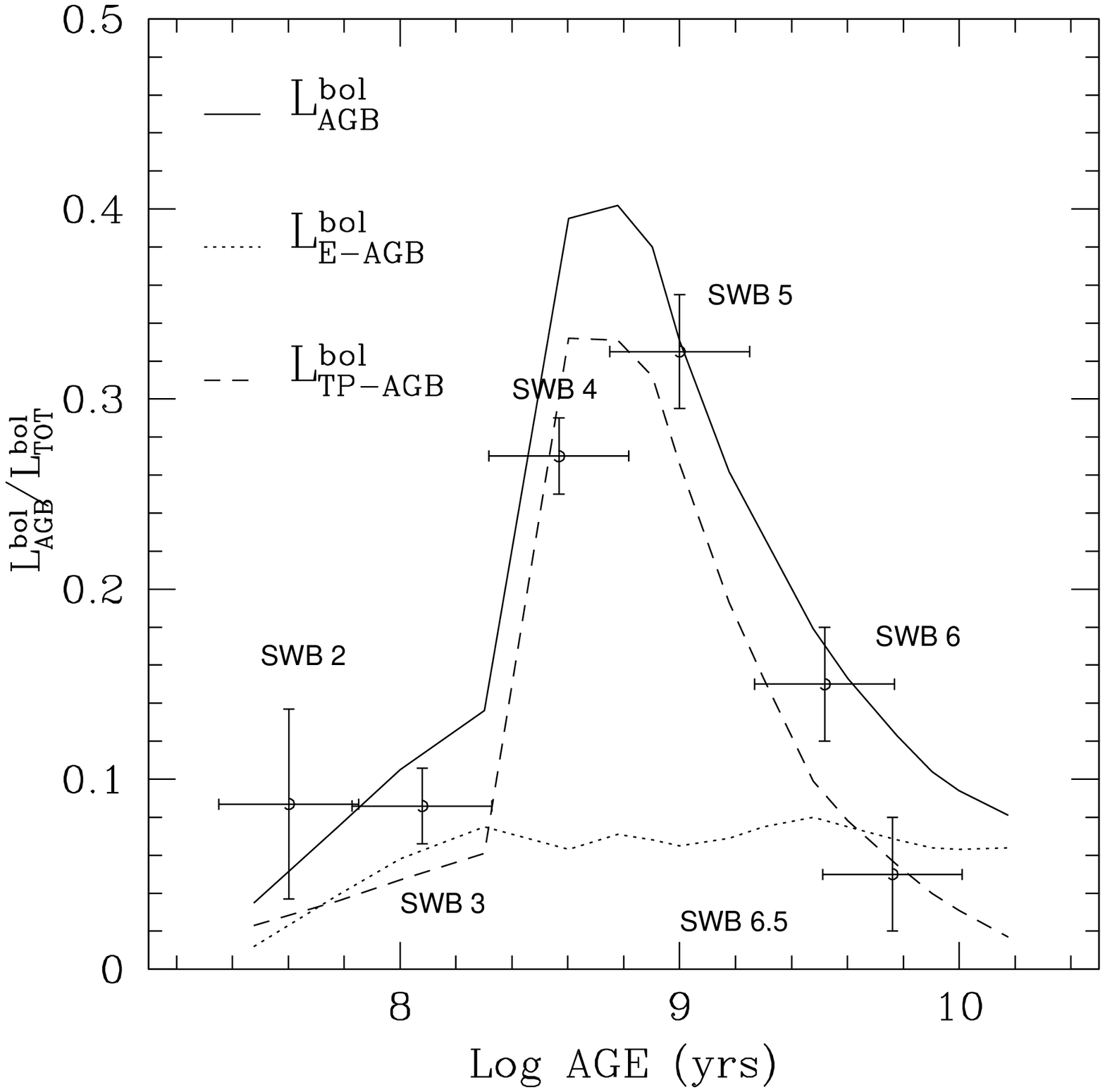}
\includegraphics[width=.49\textwidth]{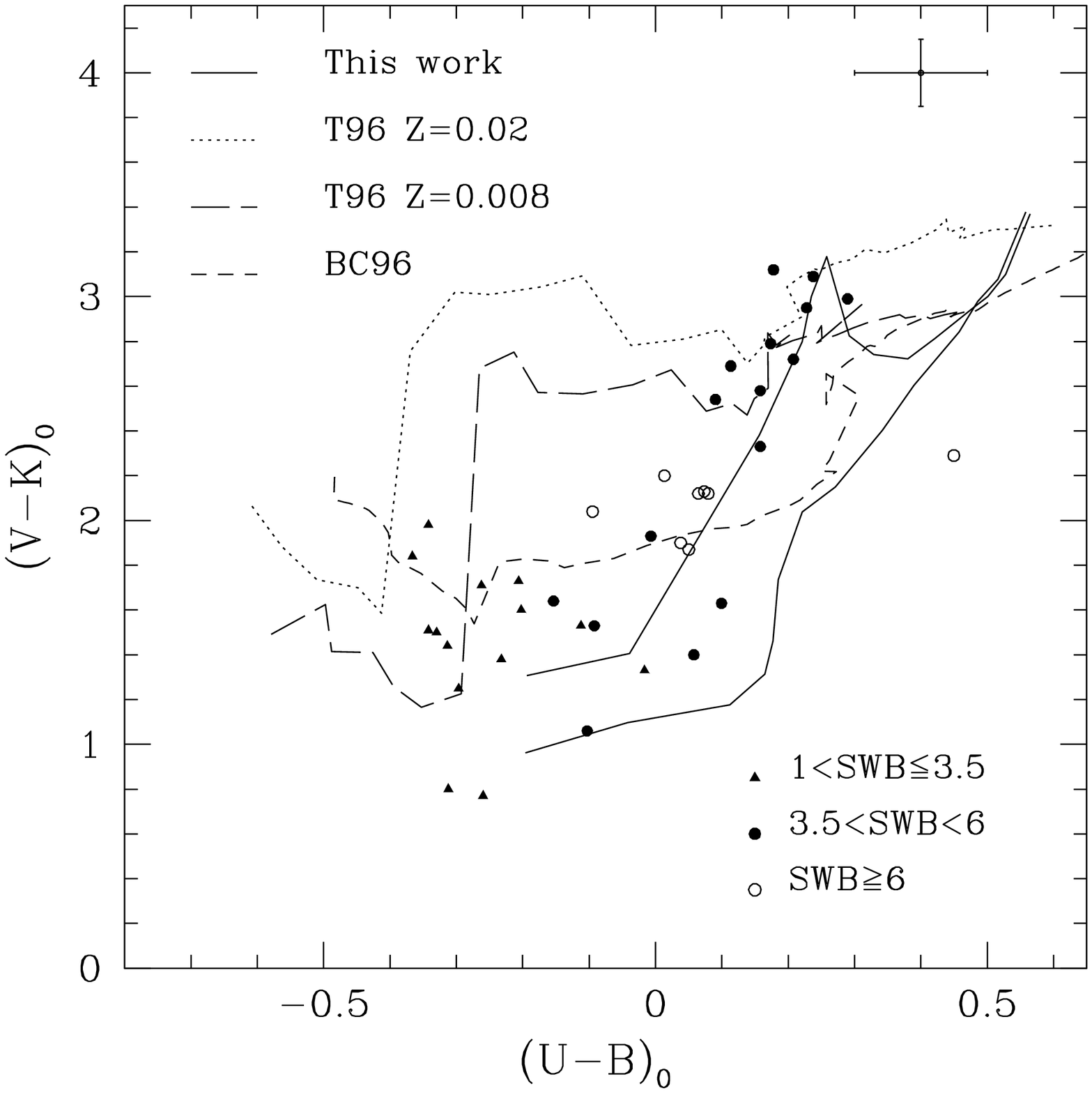}
\caption{{\it Left-hand panel}. Calibration of the bolometric
contribution of the TP-AGB phase for SSP models as a function of the
SSP age, with Magellanic Clouds GCs data. {\it Right-hand
panel}. Calibration of the SSP broad-band colours V$-$K vs. U$-$B
(solid thick line) with the same data. The filled circles are
intermediate-age GCs. The solid thin line shows the same SSPs but
without the TP-AGB phase. The other line styles show SSPs from other
authors. From \cite{io98}.}
\label{agb}
\end{figure}
This is done in SSP models in which the TP-AGB phase is calibrated
with data of intermediate-age Magellanic Clouds GCs (\cite{io98}).  It
is shown there (and in Figure~\ref{agb}) that the inclusion of this
phase is crucial to match the integrated near-IR colours of these
clusters. The calibrated models of \cite{io98} have been used to
succesfully reveal the occurrence of AGB dominated GCs in the merger
remnant galaxy NGC~7252 (\cite{ioetal01}; Figure~\ref{n7252}).

It is also shown that SSP models using uncalibrated stellar tracks for
the TP-AGB fail to reproduce the Magellanic Clouds GCs colours. As
mentioned in Section~3.4, the TP-AGB phase is included
semi-empirically also by \cite{CB91}, and the AGB bolometric
contribution is calibrated with Magellanic Clouds GCs, like in
Figure~\ref{agb}. However, the SSP integrated colours (dashed line in
Figure~3) do not exhibit the required ``jump'' in V$-$K displayed by
these GCs. This might be due to the fact that in their models the
contribution by the sole TP-AGB never exceeds $\sim~10\%$ (Figure~10
in \cite{CB91}), at variance with Figure~\ref{agb}.
\begin{figure}[ht]
\begin{center}
\includegraphics[width=.6\textwidth]{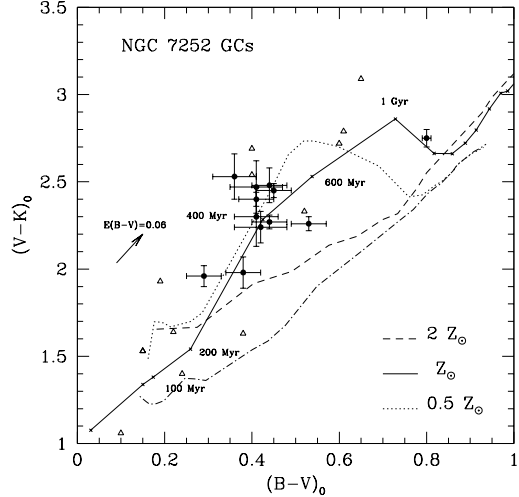}
\caption{Detection of AGB-dominated GCs in an external galaxy. Same
diagram as in Figure~\ref{agb} for the GCs of the merger remnant
galaxy NGC~7252 (filled symbols). From \cite{ioetal01}.}
\label{n7252}
\end{center}
\end{figure}
\subsubsection{The influence of nebular continua.}
The SED of very young stellar populations located in starforming
regions may be affected by emission from gas. This is what the models
of the Starburst99 group (\cite{Starburst99}) try to account
for. Specifically designed for active starforming regions, these
models include also emission lines from interstellar components. The
TP-AGB phase is not included in these models, that are therefore
appropriate for ages smaller than $\sim 100$~Myr.

\subsubsection{High-resolution SSPs}

\cite{Vaz99} presents high spectral resolution (1.8 \AA) SEDs of old
SSPs. These are obtained by using empirical stellar libraries instead
of the Kurucz model atmospheres. The advantage is that the observed
spectra can be directly compared with the models without degrading
their resolution. However, the presently very limited empirical
libraries permit to model only a rather narrow range in ages and
metallcities.

\subsubsection{Model Lick indices for non solar chemistries}

The ratio of $\alpha$-elements to Fe-peak elements (\afe) is a key
diagnostic for the formation timescales of the stellar populations in
galaxies (e.g. \cite{TGB99}). However, the standard model Lick indices
(Section~3.5) are inadequate to study stellar populations with
$\afe\neq 0$.
\begin{figure}[ht]
\begin{center}
\includegraphics[width=.49\textwidth]{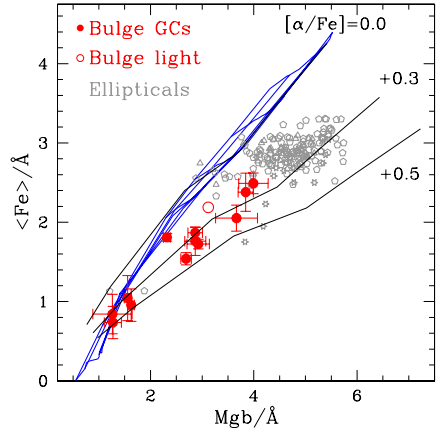}
\caption{From \cite{ioetal02}. Calibration of standard model Lick
indices (grid) with GCs including Bulge objects (filled large symbols,
from \cite{Puzia}) The three thick lines show the models by
\cite{TMB02} with a constant age of 12 Gyr and three values of the
\afe~ratios. The large empty circle is the average Bulge field in
Baade window. Small symbols are data of ellipticals from various
samples.}
\label{bulge}
\end{center}
\end{figure}
This is demonstrated by the calibration of these type of models up to
\zsun~(\cite{ioetal02}), accomplished with the metal-rich ($Z\sim
\zsun$) and $\afe=0.3$~GCs NGC~6528 and NGC~6553 of the galactic Bulge
(Figure~\ref{bulge}, \cite{ioetal02}). The model comparison with
galactic GCs with lower metallicities shows that the standard models
reflect non solar \afe~at subsolar metallicities. This {\it variable}
\afe~of the standard models results from the calibrating stars used to
compute the fitting functions (\cite{ioetal02}).  To derive the
abundances of unresolved extra-galactic stellar populations it is
desirable to have models with well-defined values of element abundance
ratios. \cite{TMB02} provide such {\it new generation} models, that
for the first time are given as functions of element abundances
(e.g. the \afe~ratio) and for various values of \afe, \aca, \an,
etc. To establish their adequacy, these models for \afe=0.3 are
calibrated with GCs (thick lines in Figure~\ref{bulge}).

\subsubsection{Horizontal Branch morphology and age confusion}

The morphology of the HB impacts on Balmer lines and broad band
colours. We focus here on Balmer lines, since they are widely used as
age indicators for GCs. The degeneracy age-HB morphology arises from
the fact that the blue HBs of metal-poor GCs enhance their integrated
\hb~values, that become larger than those of coeval GCs with red HBs.
The bluer HBs at decreasing metallicity are due to, as a first
parameter, hotter stars at lower metallicity. As a second parameter to
the mass-loss along the RGB, which by removing stellar envelope,
shifts the \teff~of the stars to higher values.  Therefore the Balmer
lines of coeval stellar populations {\it increase} with decreasing
metallicity, which undermines their power as age indicators
(\cite{MT00}).
\begin{figure}[ht]
\begin{center}
\includegraphics[width=.49\textwidth]{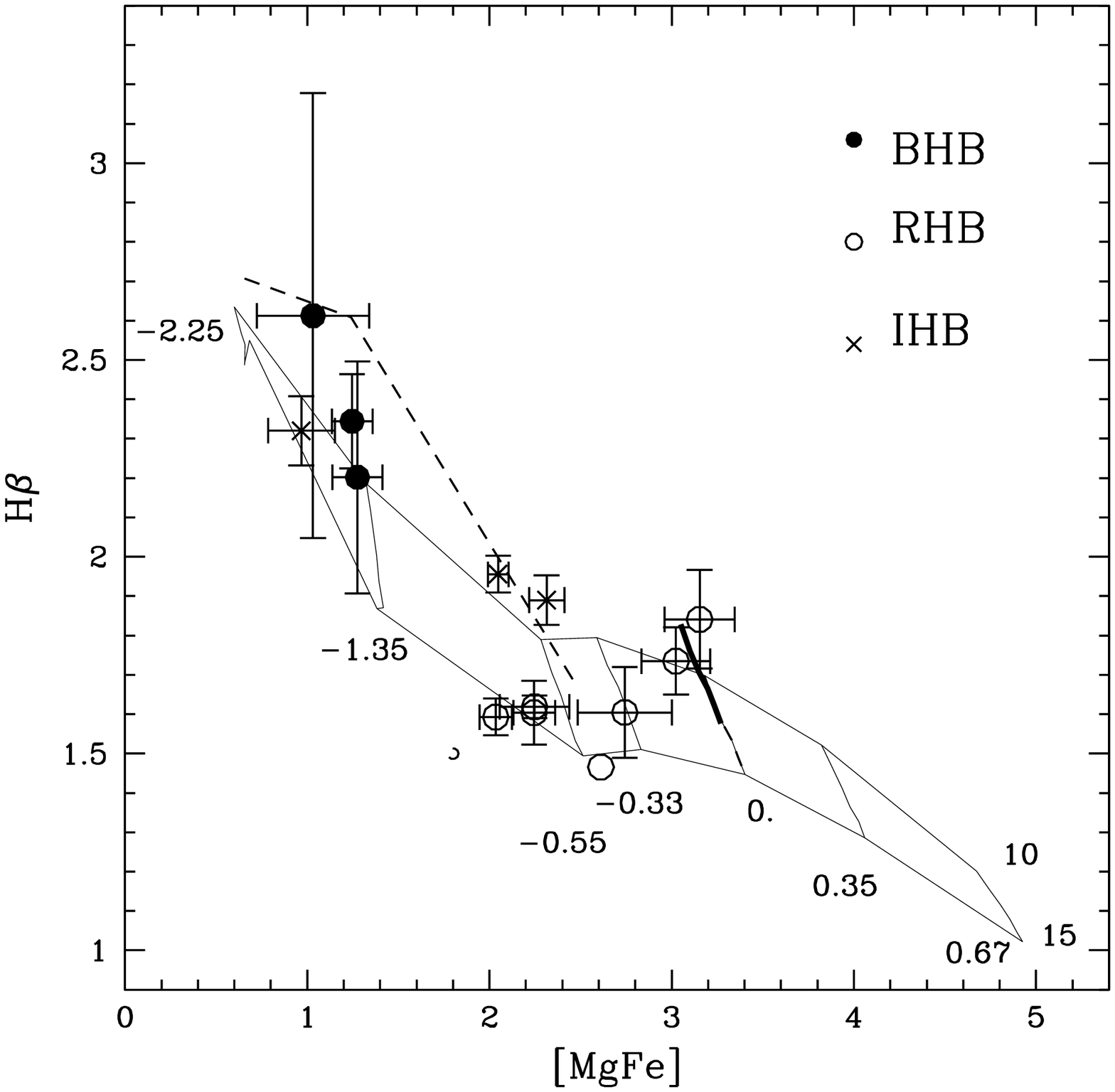}
\includegraphics[width=.49\textwidth]{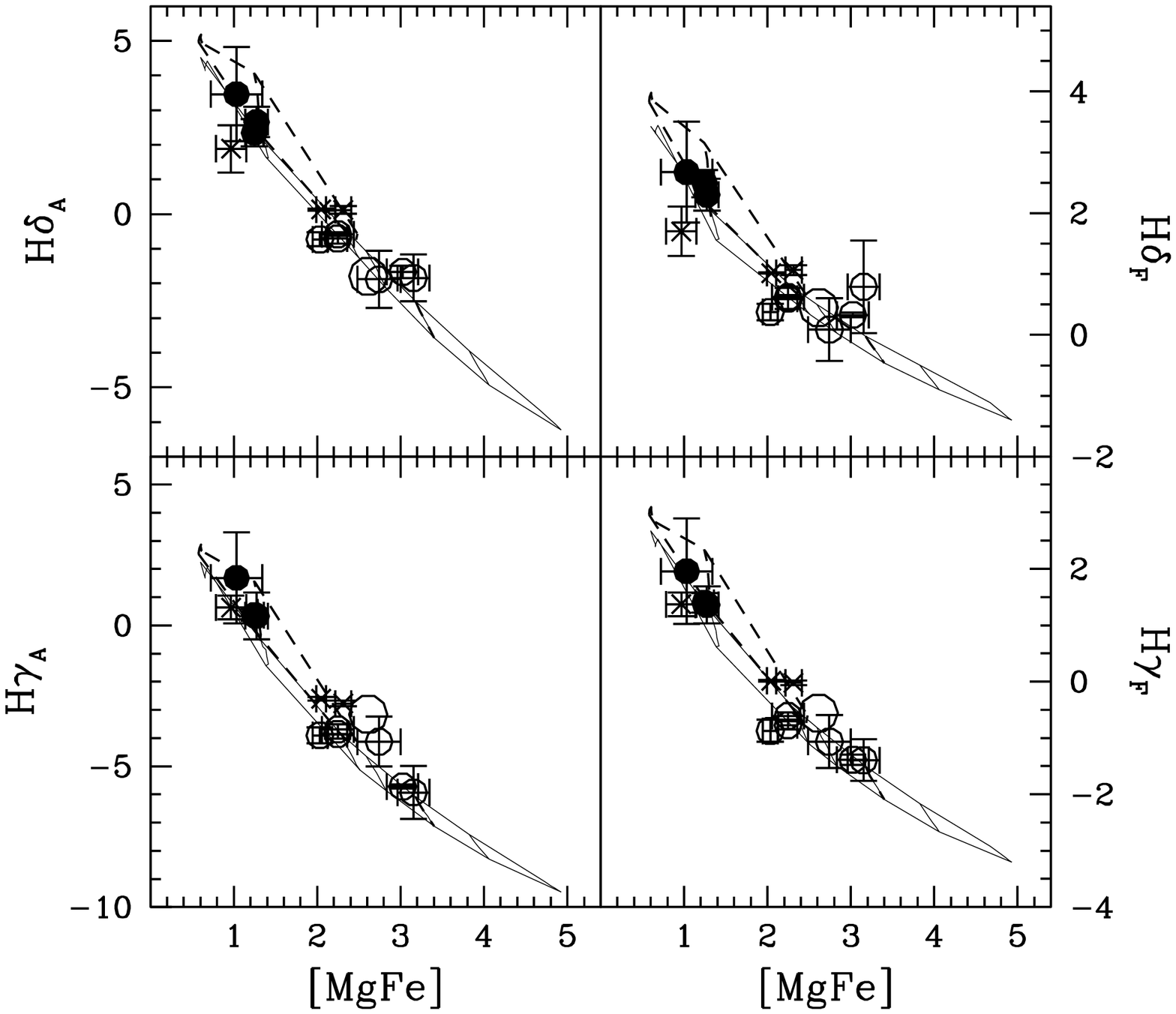}
\caption{Calibration of Balmer lines vs. HB morphology with galactic
GCs, from \cite{ioetal02}. Left-hand panel: \hb; right-hand panel:
\hda, \hga, \hdf~and \hgf. HB morphologies of GCs are coded, in which
RHB, BHB and IHB refer to red, blue and intermediate HBs,
respectively. Dashed line: 15 Gyr old SSPs with various \zh~(labelled)
and canonical mass-loss. Solid lines: 15 and 10 Gyr old models (for
same \zh) in which no mass loss has been assumed, in order to mimic
reddish HB morphologies.}
\label{hb}
\end{center}
\end{figure}
As mentioned in Section~2.2, the assumed mass loss in SSP models must
be calibrated a posteriori. \cite{MT00} calibrate the mass-loss
required to reproduce the \hb~line observed in galactic GCs as a
function of metallicity. An updated version of that calibration, which
extends to higher-order Balmer lines with the aid of new data is shown
in Figure~\ref{hb} (from \cite{ioetal02}).  The age ambiguity induced
by the HB morphology is nicely illustrated by the four GCs at
$\zh\sim-0.6$. Two of them (NGC~6441 and NGC~6338) have blue HB stars,
which are responsible for their rather large \hb's ($\sim~2$
\AA). Note that if this effect is not accounted for, these two GCs
would appear considerably younger ($\sim~8$~Gyr) than those with same
\zh~that lie on the 15 Gyr old SSP, in spite of the very similar ages
derived from their MS turnoff.

\end{document}